# Using Linked Micromaps for Evidence-Based Policy

Randall Powers[1], John Eltinge[2], Wendy Martinez[2], Darcy Steeg Morris[2]

**Abstract**:

Linked micromaps were originally developed to display geographically indexed statistics in an intuitive way by connecting them to a sequence of small maps. The approach integrates several visualization design principles, such as small multiples, discrete color indexing, and ordering. Linked micromaps allow for other types of data displays that are connected to and conditional on geographic areas. Initial applications of micromaps used data from the National Cancer Institute and the Environmental Protection Agency. In this paper, we will show how linked micromaps can be used to better understand and explore relationships and distributions of statistics linked to US states and Washington, DC. We will compare linked micromaps with other popular data displays of geographic data, such as bubble maps, choropleth maps, and bar charts. We will illustrate how linked micromaps can be used for evidence-based decision-making using data from the Bureau of Labor Statistics, the Census Bureau, and the Economic Research Service. The presentations, R scripts, and the data sets used in this article are available here: https://github.com/wlmcensus/Joint-Statistical-Meetings-Presentation-2024. The work discussed in this article was presented at the Joint Statistical Meetings (JSM) 2024 and the American Association for Public Opinion Research (AAPOR) 2024 Annual Conference.

**Acknowledgements and Disclaimer**: The authors thank Rose Byrne, Jeffrey Gonzalez, Evan Moffett, and Emily Molfino for their helpful comments. The views expressed in this article are those of the authors and do not represent the policies of the Bureau of Labor Statistics and the United States Census Bureau. All empirical results included in this article depend only on data previously cleared for dissemination and are provided through public-access websites.

## 1. Introduction

The goal of this work is to review a visualization approach called linked micromaps and to highlight some ways in which that approach can be useful for the exploration of policy-relevant statistical information. This paper also discusses some ways in which linked micromaps can be better for visualizing, exploring, and understanding spatially indexed data than the ways typically used to display such data. The idea for linked micromaps was originally developed by Carr and Pierson (1996) in response to dissatisfaction with how choropleth maps were used to understand economic data from the Bureau of Labor Statistics (BLS) while Dan Carr was there as a Research Fellow. He

---

[1] Corresponding author, Bureau of Labor Statistics, powers.randall@bls.gov

[2] U.S. Census Bureau

was charged with developing "new and innovative methods for displaying the data." (Carr and Pierson, 1996)

Linked micromaps provide visual summaries of statistics, data, and geographic distributions in a coherent and intuitive manner. They provide accessible tools for exploring and disseminating statistical information–including complex distributional and multivariate patterns–connected with geography. In general, linked micromaps consist of several columns, with one column depicting a set of small maps and areal or polygonal units that are linked across rows to other columns showing statistics of interest. Details on the components of linked micromaps will be covered in a later section; see Figure 4 for an example of a linked micromaps graphic.

While original tools for linked micromaps used Java (Carr and Pickle, 2010), statisticians and data scientists can easily create them now using two R packages – `micromap` and `micromapST`– which are available on CRAN (Payton, et al., 2015; Pickle, et al., 2015). The linked micromap visualizations can be saved to a graphics file and imported into documents and presentations.

This paper uses the `micromapST` package to describe the components of linked micromaps and for the applications in Section 4. The `micromapST` package was created as an easy-to-use interface for creating linked micromaps where the geographies of interest are the 50 states and Washington, DC. Such geographic data are typically of interest in US federal statistical agencies. The next section of the paper describes some of the common ways used for visualizing spatially indexed statistics and their limitations. This is followed by a more detailed description of the components in a linked micromap. The paper concludes with applications of linked micromap using public-domain data from the following sources:

- The American Community Survey (ACS, https://www.census.gov/programs-surveys/acs)
- The Quarterly Census of Employment and Wages (QCEW, https://www.bls.gov/cew/)
- The Economic Research Agency Food Atlas (https://www.ers.usda.gov/data-products/food-access-research-atlas/).

## 2. Visualizing Spatially Indexed Data

Before exploring the features and usefulness of linked micromaps, it is helpful to provide some context by reviewing other methods of displaying state-level data. Examples of typical methods described in this section include bar charts, tables, bubble maps (also called proportional symbol maps), and choropleth maps.

Bar charts are a common way to display state-level statistics, in part because they are easily created in Microsoft Excel and most statistical programs. Horizontal bars are typically used where the length of each bar represents the statistic of interest (e.g., total number of employees in a state, etc.). Figure 1 displays a bar chart from Excel showing the ACS Response Rates for 2022. The bars are ordered alphabetically by state. Within this specific Excel implementation, the default axes labelling for the size of the chart is such that every other bar is labeled, so we lose half of the state names. Because of these issues, it is difficult to extract useful information about the statistics as they relate to geographies. One could also order the data by the statistic, which provides information on their distribution. However, there is still a problem of understanding spatial distributions and their connections with the statistics.

State-level data are often presented in a table and again usually ordered alphabetically by state. As with bar charts, this method of data presentation makes it difficult to extract information and insights regarding the connection between the statistic and geography. It is up to the viewer to make a mental comparison of data throughout the table.

Choropleth maps are often used to display geographic data connected to an areal unit, such as a state or county. In choropleth maps, the color indicates some variable of interest; see Figures 2 and 11 for examples. Choropleth maps are very popular for good reasons. They are intuitive for most people to understand, and they show spatial distributions of one characteristic of data quite well. However, there are some limitations (Carr and Pierson, 1996). These include the following:

- It is sometimes difficult to distinguish values using color, especially when color gradients are used.
- Additional information related to the statistics of interest is hard to convey on a static map.
- There is no ordering of the statistics (and hence the geography), except through the legend.
- Choropleth maps are most appropriate for proportions or other normalized data, not raw counts.
- As given by their nature, choropleth maps give the less populated but large land mass states more prominence, making less populous states visually more distinct or important and tiny states like Rhode Island easy to miss.

An enhancement to choropleth maps is the bubble or proportional symbol map. This approach adds bubbles or circles over each region that represents one or two additional values using the size of the bubble and the color; see Figure 3 for an example. These also have limitations (Robbins, 2013), which include:

- A bubble map can be difficult to understand when comparing two or more statistics per areal unit (e.g., color of region, color of bubble, size of bubble).
- The overlapping of bubbles can make it difficult to distinguish them.
- It can be hard (impossible?) to ascertain exact values in static proportional symbol maps using circle sizes & color scales.

## 3. Linked Micromaps

Building on the context provided by Sections 1 and 2, the remainder of this article is focused on the graphical displays from the R `micromapST` package, which is used for visualizing statistics linked to the 50 US states and Washington, DC. The `micromapST` graphics follow the main elements of linked micromaps, which are described next.

Linked micromaps consist of at least three vertical columns containing (1) a color legend indicating each areal unit, (2) a series of small maps of the overall geographical area (e.g., the US) with colored or highlighted regions indicating the current row (e.g., a state), and (3) a statistic (e.g., response rate). See Figure 4 for an example of a linked micromap showing 2022 unit-level ACS household response rates for each state. Note that the statistic of interest is ordered (response rate), and that the color of the regions is meant to connect the statistic across the rows to the region.

The regions are organized into perceptual groups (using the default in `micromapST`) to make it easier to understand the graphic and to extract useful information. The US regions are readily divided in a rather pleasing and intuitive way. First, they are divided according to the median value of the ordered statistic, with 25 regions above and 25 below. Those are further subdivided into groups of five. Note that the state with the median sorted variable is in its own group. Showing just five states in each map/group helps the viewer more easily identify the states, especially the small ones like Rhode Island or Washington, DC that are easier to miss in a single choropleth of all states.

Figure 4 shows just one type of plot for a statistic, which is a dot plot. The `micromapST` package allows for the following types of graphics: arrows, bars, boxplots, dots, scatter plots, and time series. Some examples of these will be given in Section 4. Furthermore, more than one statistic (i.e., column of data or statistics) can be shown in the linked micromaps graphic. More information on the package, available plot types, glyphs, examples, and other options (e.g., conveying uncertainty) can be found in Pickle, et al. (2015).

## 4. Applications

### 4.1 American Community Survey Response Rates

The American Community Survey (ACS) is an annual survey that collects information on the nation and its people, generating data on demographics, economics, housing, and more. Unfortunately, response rates for the ACS have been falling, as is the case with many government surveys today. Thus, survey managers have interest in understanding factors that may be associated with , which may point toward possible mitigation options. Exploring the unit-level ACS household response rates using linked micromaps can provide potentially useful insights for this purpose.

Unit-level ACS household response rates for each state and over time are publicly available on the following US Census Bureau site:

www.census.gov/acs/www/methodology/sample-size-and-data-quality/response-rates/

One can view and download response rates for the nation as a whole or by states from 2000 to 2023. Response rates and reasons for non-interviews are displayed on the website in tabular form. One can either see the national response rates over time (temporal distribution of response rates for the nation as a whole) or the response rates for each state in a single year (spatial distribution for a given year). In other words, the commonly used tables show the marginal distribution over time or space but do not readily support exploration of the joint distribution of these rates over both space and time.

This shows some of the limitations of using tables and where linked micromaps can help display information on multiple aspects of the data. Figures 4 and 5 show a linked micromap depicting the unit-level ACS household response rates for a single year–2022. It seems odd that the highest response rates are for Utah and Idaho, while the lowest response rate is for Washington, DC. While there are many factors at play that might affect response rates (e.g., demographics, economics), one might expect a region like Washington, DC containing primarily government activities would have a higher response rate to a government survey.

Another relevant question for survey managers may be: "How do the response rates vary over time and space?" One of the plot types in the `micromapST` package is a time series. A column showing the response rates from 2010 through 2022 was added to the linked micromaps plot (see Figure 6). The time series plots help convey information over space (the states) and time. Two interesting years are immediately apparent. There is an obvious dip in response rates in 2013, which was mostly due to a two-week government shutdown. A more extreme dip in response rates happened in 2020,[3] which was mostly due to COVID-19 quarantines, lockdowns, and other data collection interruptions in ACS operations. It is interesting to note that the states suffering the lowest COVID-19 drops in response rates also displayed relatively slow trajectories of recovery.

Survey managers may also want to understand the relationship between certain characteristics or beliefs and potential trends in response rates. For example, a possible reason for not responding to the ACS is anti-government sentiment. The Pew Research Center conducted the Religious Landscape Study in 2014.[4] One of the questions asked respondents: "If you had to choose, would you rather have a smaller government providing fewer services, or a bigger government providing more services?"; i.e., did they like big government or small government? The percent pro small government can be used as a potential indicator of anti-government sentiment. To explore this, Figure 7 linked micromaps that include the state level "pro small government" proportions from the 2014 Pew Research study, along with the 2022 state-level unit-level ACS household response rates, and the state-level declines in response rates between 2010 and 2022. The linked micromaps do not display an obvious relationship between the response rate patterns and the Pew Research sentiment statistics. In addtion, Washington, DC stands out as an outlier, in that it has the lowest response rate and the lowest percentage pro small government (meaning it is generally pro large government), with both values somewhat different from other states.

---
[3] https://www.census.gov/library/working-papers/2021/acs/2021_CensusBureau_01.html

**4.2 Quarterly Census of Employment and Wages – During the Early COVID-19 Years**

The next example of linked micromaps uses data from the Bureau of Labor Statistics (BLS). The Quarterly Census of Employment and Wages (QCEW) publishes state level data on the BLS website and is publicly available for download. The QCEW has quarterly state and county level data for numerous industries, for both employment and wages, as well as ownership (private or government). After looking at the unit-level ACS household response rates over time, it seemed that there might be a story to tell regarding the COVID-19 pandemic and employment. How did reported employment change during the early years of the pandemic? Did those changes show different patterns across different industries? Were the patterns of change similar across states? After looking at several industries, a decision was made to focus on the Leisure and Hospitality industry, which displayed notable decreases in employment near the beginning of the pandemic in March 2020.

An interactive tool (https://data.bls.gov/maps/cew/us) for QCEW state level data allows a user to choose an industry and specific series variable to extract. The results can then be rendered in a choropleth map and summary table. At the bottom of the page is a link to download the data shown in the map.

As discussed in a previous section, choropleth maps have some limitations. For example, we are interested in how the employment statistics change over time. Using choropleth maps, we would need to show a series of choropleth maps at different time points. Figure 8 shows such a series of maps obtained using the QCEW tool, where four choropleth maps at different points in time are displayed, each one showing the 12-month percent change in employment for the Leisure and Hospitality industry. Readers can generate these maps easily on the QCEW site.

The map on the upper left is for Quarter 4 of 2019 (12-month percent change from December 2018 to December 2019), right before the pandemic and subsequent quarantines. Note that the blueish colors represent a positive change over the most recent 12-month period. The yellow and orangey colors represent a decrease in employment over the most recent 12-month period. From the colors, it can be observed that employment numbers were on the rise for the vast majority of the states in Quarter 4 of 2019.

The map on the upper right is Quarter 1 of 2020 (March). Recall that this is when the pandemic hit, and widespread closures ensued. An immediate downturn in the employment number in Leisure and Hospitality for almost all states can be seen by the large number of yellow and orange states, showing an immediate effect on employment in this industry. The lower left is a year later for Quarter 1 of 2021 (12-month percent change from March 2020 to March 2021), and now all states are showing a negative change (all yellow-orange colors). However, the story changes only one quarter later (Quarter 2 of 2021) with recovery in employment beginning in many states, roughly coincident with the widespread introduction of COIVD-19 vaccines (Watson, et al., 2022).

---

[4] https://www.pewresearch.org/dataset/pew-research-center-2014-u-s-religious-landscape-study/

It is difficult to see in these plots (readers can generate these maps easily on the QCEW site), but the color scales for each of these maps is different, making it hard to compare actual quantitative differences across the maps from color alone. For example, dark orange does not have the same numerical range across the maps. We only know it corresponds to the largest negative change for that quarter and year. Also, this graphic represents four different snapshots in time from the webpage that have been copied and pasted together, and resolution is poor. It would be preferable to be able to display the data in a true time series fashion where the relevant statistics are presented with a single legend to allow for true comparisons. A story of the effects of the pandemic on the Leisure and Hospitality industry employment situation can be told using choropleth maps, but it is not easily done.

The same data will now be visualized using the `micromapST` package. In Figure 9, the states have been sorted based on the 2020 Q1 Over-the-Year Change for the Leisure and Hospitality industry. This clearly shows similar information to the choropleth maps – that the downturn in Leisure and Hospitality jobs was almost immediate across the US. However, the linked micromaps provide information on the values (no longer referencing an interval) and magnitudes of the change. For example, we see that Idaho, Wyoming, and Montana were the top three states showing an increased percentage of reported employment in Leisure and Hospitality.

We highlighted some interesting features or patterns in the same linked micromaps as shown in Figure 10. The first statistical column shows a time series of the over-the-year change from 2020 to 2022. The last data column displays arrows indicating the difference in one-year employment from the start of the pandemic to recovery in 2022 (Q1). Some interesting behaviors that policymakers might want to explore are highlighted (circled) in the figure. For example, Arizona had some extreme swings in employment over the period, as seen in the time series column, and Washington, DC had the largest percent change in employment. Linked micromaps give the user the ability to explore and gain insights on how the pandemic affected employment across the nation and through time.

**4.3 Economic Research Service Food Atlas**

The next example is taken from the Economic Research Service's (ERS) Food Environment Atlas website www.ers.usda.gov/data-products/food-environment-atlas/go-to-the-atlas/. The objectives of the atlas are:

- "To assemble statistics on food environment indicators to stimulate research on the determinants of food choices and diet quality, and
- To provide a spatial overview of a community's ability to access healthy food and its success in doing so."[5]

The ERS Food Environment Atlas features a plethora of variables, each of which the user can choose for the production of a choropleth map of the data, one variable at a time. Some variables feature data from one year, some of them offer an average over a certain number of years, some of them show the change over time, and so forth. Some are at the state level, and some are at the county level. While this atlas offers a wealth of information, it is arguably difficult to use choropleth maps to explore relationships among the variables, because each choropleth map from the Atlas is limited to one variable. For example, Figure 11 depicts two maps: a state-level choropleth map showing food insecurity (percent three-year average), and another map shows population totals with low access to stores at the county level. If the user wanted to explore the relationship between two or more variables, then they must compare two or more choropleth maps. Recall from the QCEW example that this can be difficult to do.

---

[5] https://www.ers.usda.gov/data-products/food-environment-atlas/

ERS data used in the Food Environment Atlas was then explored with the help of linked micromaps. The mission of the ERS is to "anticipate trends and emerging issues in agriculture, food, the environment, and rural America and to conduct high-quality, objective economic research to inform and enhance public and private decision making."[6] Using data as evidence for making and assessing policy is critical, and finding relationships and interesting phenomena (hot spots, extremes, etc.) through data visualizations is a key step in understanding situations.

One government food assistance program is SNAP (Supplemental Nutrition Assistance Program). Simply put, this program puts food dollars in the hands of people who need it to improve their health by eating nutritious food. Food insecurity is the condition of not having access to sufficient food, or food of an adequate quality, to meet one's basic needs. Policymakers might ask "How might changes in SNAP participation impact Food Insecurity?" We can explore potential relationships between SNAP participation and food insecurity using linked micromaps, as shown in Figure 12, which displays the percent change in SNAP participation from 2012 to 2017 as dots. The change in 3-year average in household food insecurity is shown as bars in the last column. Food insecurity for most states decreased, except for New York, Nevada, Pennsylvania, and Maine. Policymakers might want to look closely at what happened in New York, since that state had the largest increase in food insecurity, as well as Nevada and Pennsylvania since those two states had positive changes in food insecurity, i.e., there were more food insecure households, despite more households participating in SNAP.

Another linked micromap showing ERS data is given in Figure 13. This graphic shows some of the other glyphs one can use with linked micromaps: boxplots and scatterplots. Policymakers might ask "Is there a relationship between Low Access to Food Stores and Food Insecurity? If Access to Stores is improved, might that affect Food Insecurity?" The sort variable in this plot is the change in Food Insecurity between two time periods. The first column shows boxplots of the county level Percent Change in Low Access to Stores from 2010 to 2015. These boxplots help users understand the distributions of the statistic over the counties (i.e., skewed, symmetric), the dispersion, and a measure of location (the median). The last column shows a scatterplot of 2015 Percent Food Insecure as a function of the 2015 Percent of Households with Low Access to Stores. There does not seem to be a relationship between them. However, one can see some interesting states highlighted with circles. The statistics for Alaska and Washington, DC seem to be different from the rest of the nation, which is readily apparent in linked micromaps. They are both extreme points in the scatterplots for access to stores, and the boxplot for Alaska is highly skewed. Policymakers might want to look more closely at these areas for additional insights.

## 5. Discussion

Linked micromaps are a way to summarize statistical information tied to spatial areas, such as countries, states, counties, etc. The R `micromap` package (Payton, et al., 2015) provides functions for producing linked micromaps for a given geography or spatial areal unit. The `micromapST` R package (Pickle, et al., 2015) was specifically written for data associated with the 51 US states and was the focus of this work. In this article, we showed how linked micromaps address many of the limitations found in the usual ways of displaying such data (e.g., bar plots, choropleth maps) through several real-world examples. The reader is encouraged to consult the open-access article by Pickle, et al. (2015) for more details on using the `micromapST` package and the many ways it can be used to explore US state-level statistics.

---

[6] https://www.ers.usda.gov/about-ers/

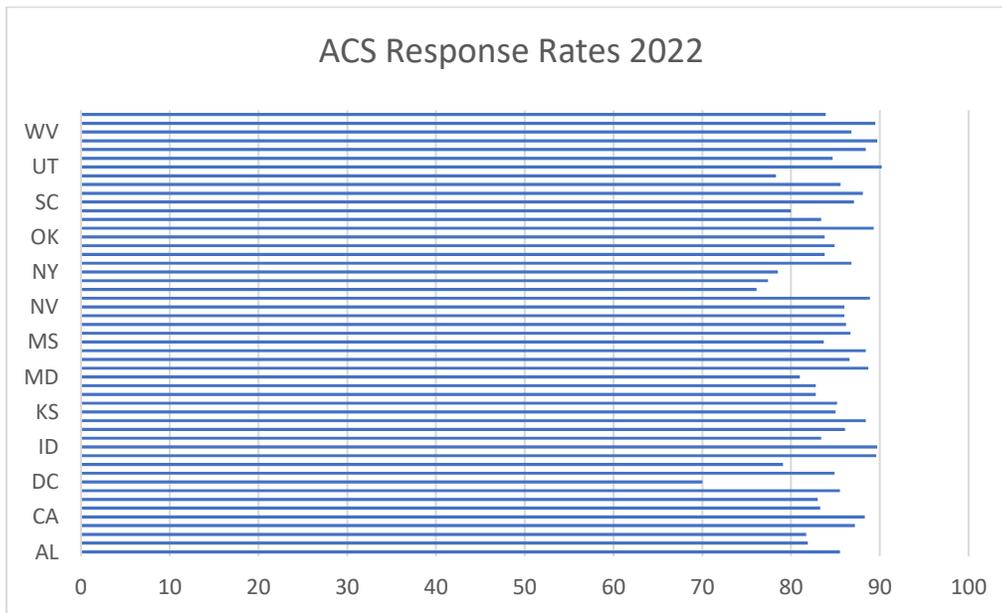

*Figure 1. Default bar chart in MS Excel of the ACS response rates for 2022 where the bars are ordered alphabetically by state. Note that only every other bar is labeled, and it is difficult to understand the distribution of the response rates over the states. SOURCE: American Community Survey, US Census Bureau, https://www.census.gov/acs/www/methodology/sample-size-and-data-quality/response-rates/.*

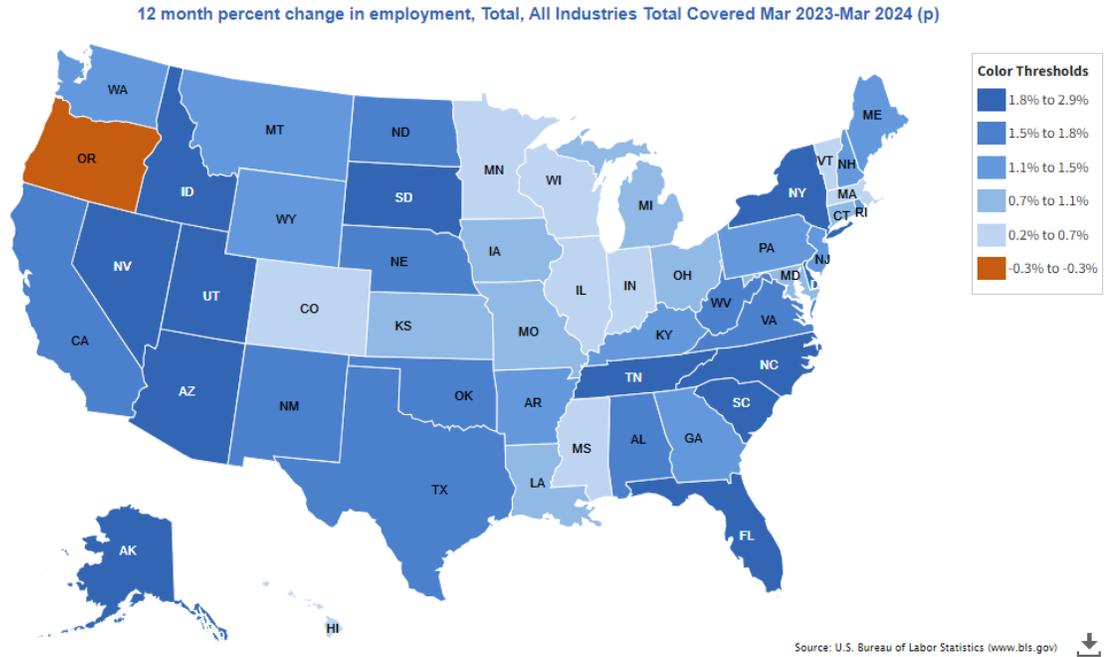

*Figure 2. Example of a choropleth map where the color corresponds to the value of a statistic (total employment). SOURCE: Quarterly Census of Employment and Wages, Bureau of Labor Statistics, https://data.bls.gov/maps/cew/us.*

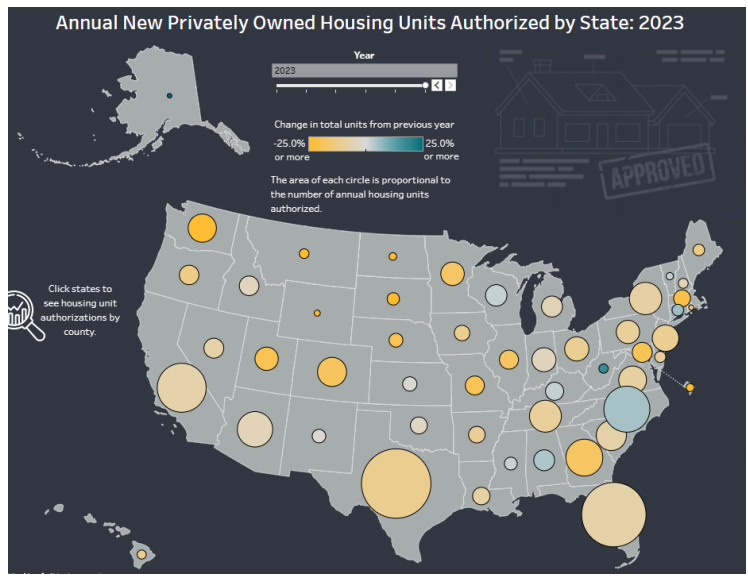

*Figure 3. This is an example of a bubble map, where the size of the bubbles and color represent statistics of interest. This is a static version of the interactive map that can be viewed at the link provided. SOURCE: US Census Bureau https://www.census.gov/library/visualizations/interactive/annual-new-privately-owned-housing-units.html*

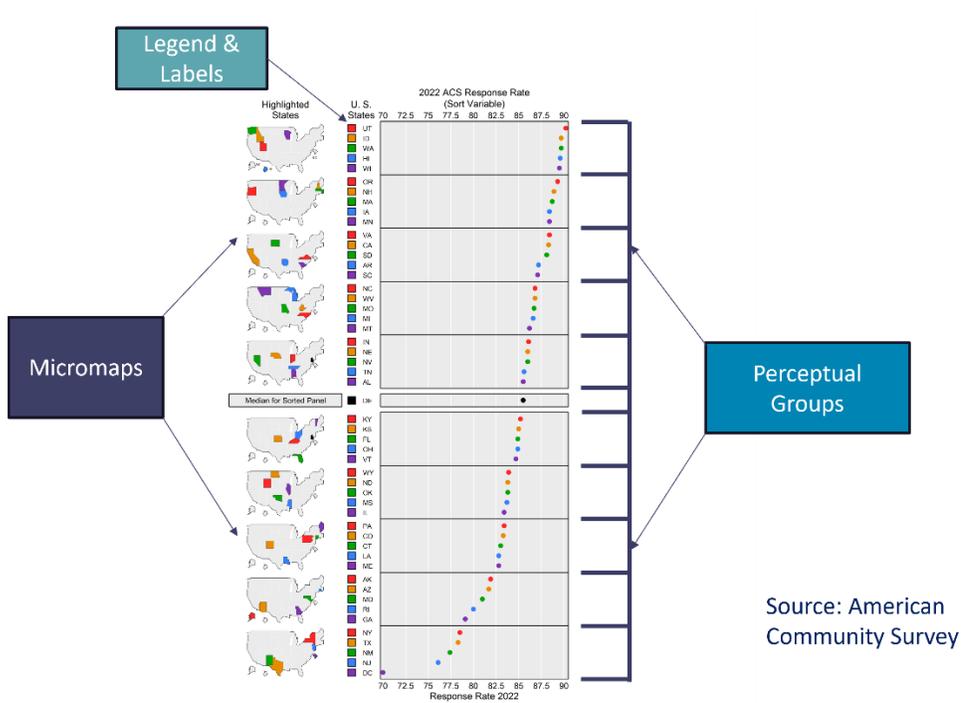

*Figure 4 This is an example of linked micromaps created using the `micromapST` package. It shows the ACS 2022 response rates for each state and Washington, DC. The general elements of linked micromaps are identified. SOURCE: American Community Survey, US Census Bureau [https://www.census.gov/programs-surveys/acs](https://www.census.gov/programs-surveys/acs).*

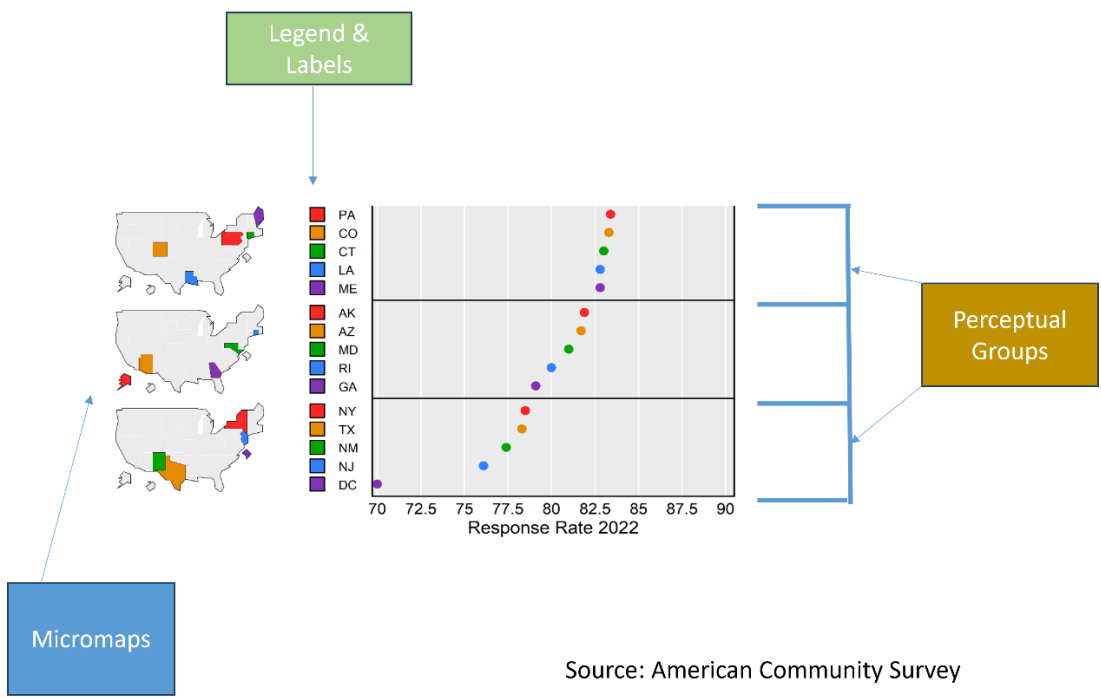

*Figure 5. This shows the three groups with the lowest response rates for the 2022 ACS, taken from the linked micromaps in Figure 4. These areas may be of special interest for further investigation.*

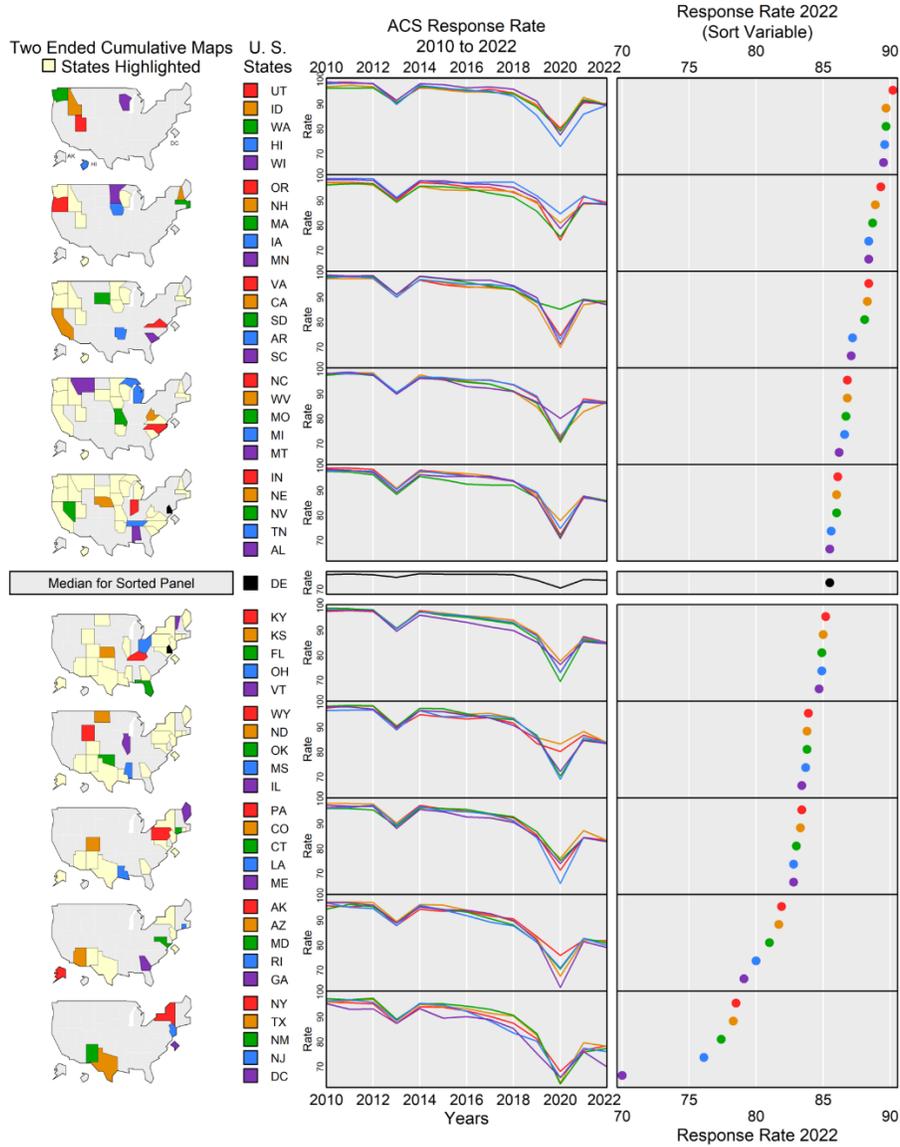

*Figure 6. This linked micromaps graphic has an additional column—a unit-level ACS household response rate time series for each state from 2010 to 2022. Two interesting years are clearly visible: the dip in rates for 2013, in which a two-week government shutdown occurred, and 2020 showing the impact from the COVID-19 pandemic. Note the different way to depict the sub-regions in the small maps (Pickle, et al., 2015) showing cumulative information for the ordered statistic. SOURCE: American Community Survey, US Census Bureau https://www.census.gov/programs-surveys/acs.*

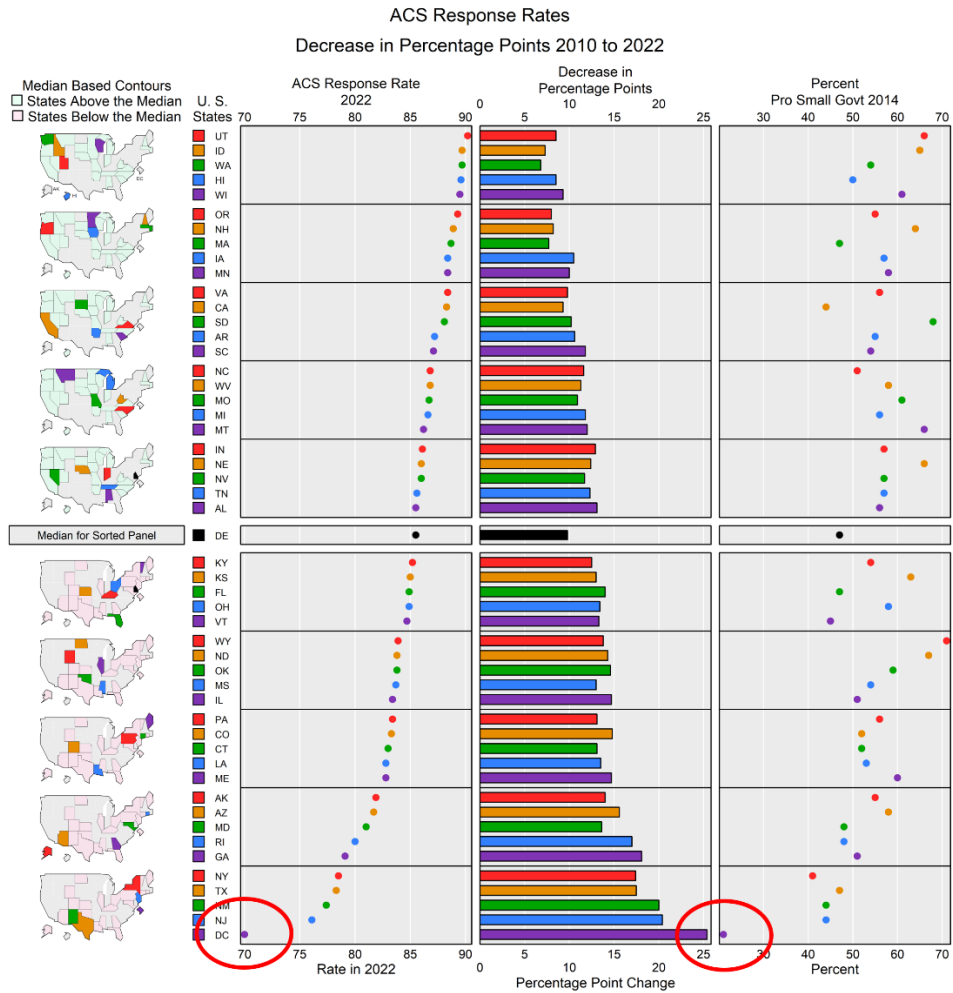

Figure 7. Here is a linked micromaps graphic with an additional column of bars indicating the decrease the unit-level ACS household rate in percentage points between 2010 and 2022. The final column presents state-level attitudes toward government as reflected in Pew Research Center (2014). There does not seem to be a relationship between the response rates and the attitudinal measure (first and third columns with summary statistics).  Washington, DC is interesting. It has the lowest response rate and the smallest attitude measure. Note the different way to depict the cumulative information connected to the sub-regions in the small maps (Pickle, et al., 2015). SOURCE: American Community Survey, US Census Bureau www.census.gov, Pew Research Center https://www.pewresearch.org.

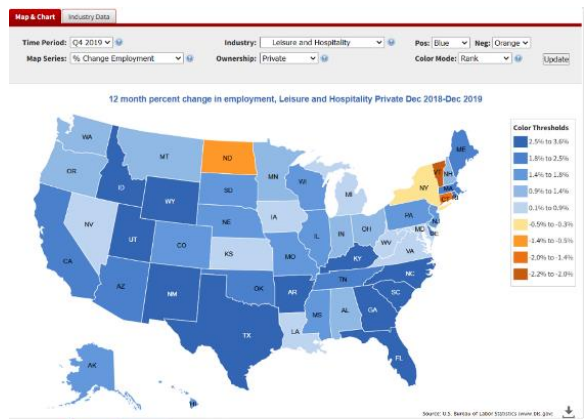
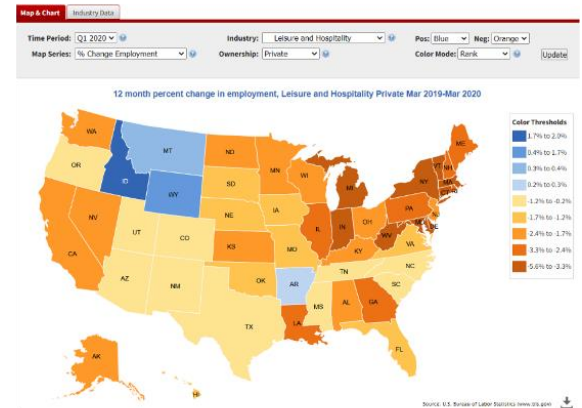
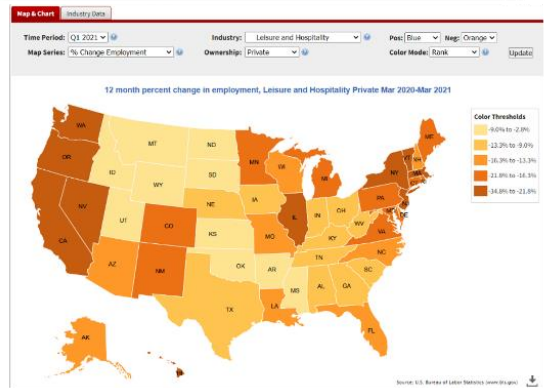
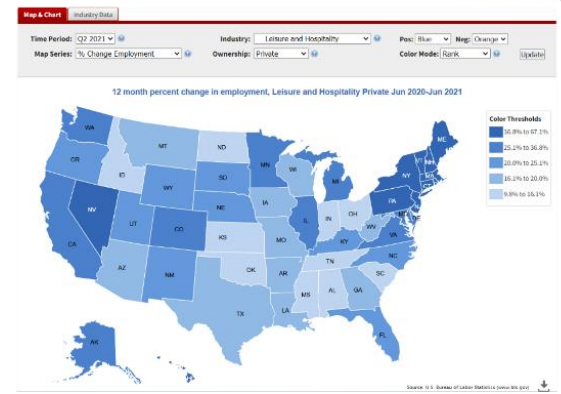

*Figure 8. This is a series of choropleth maps showing the over-the-year change in employment in the Leisure and Hospitality Industry, as measured in the QCEW. The upper left is Q4 2019; the upper right is Q1 2020 (when the impact of COVID-19 started); the bottom left is Q1 2021 (one year later); and the bottom right is Q2 2021 (one quarter later). The yellow-orange values indicate decreases in employment, and the blue represents increases in employment. One can only determine increases and decreases by the colors, not actual values. Further complicating the comparisons across time, the color thresholds (intervals) are different for each map. SOURCE: Bureau of Labor Statistics, QCEW [www.bls.gov](www.bls.gov).*

# Effects of COVID: QCEW % Change in One-Year Employment
## Leasure & Hospitatlity 2020 Q1 to 2022 Q1

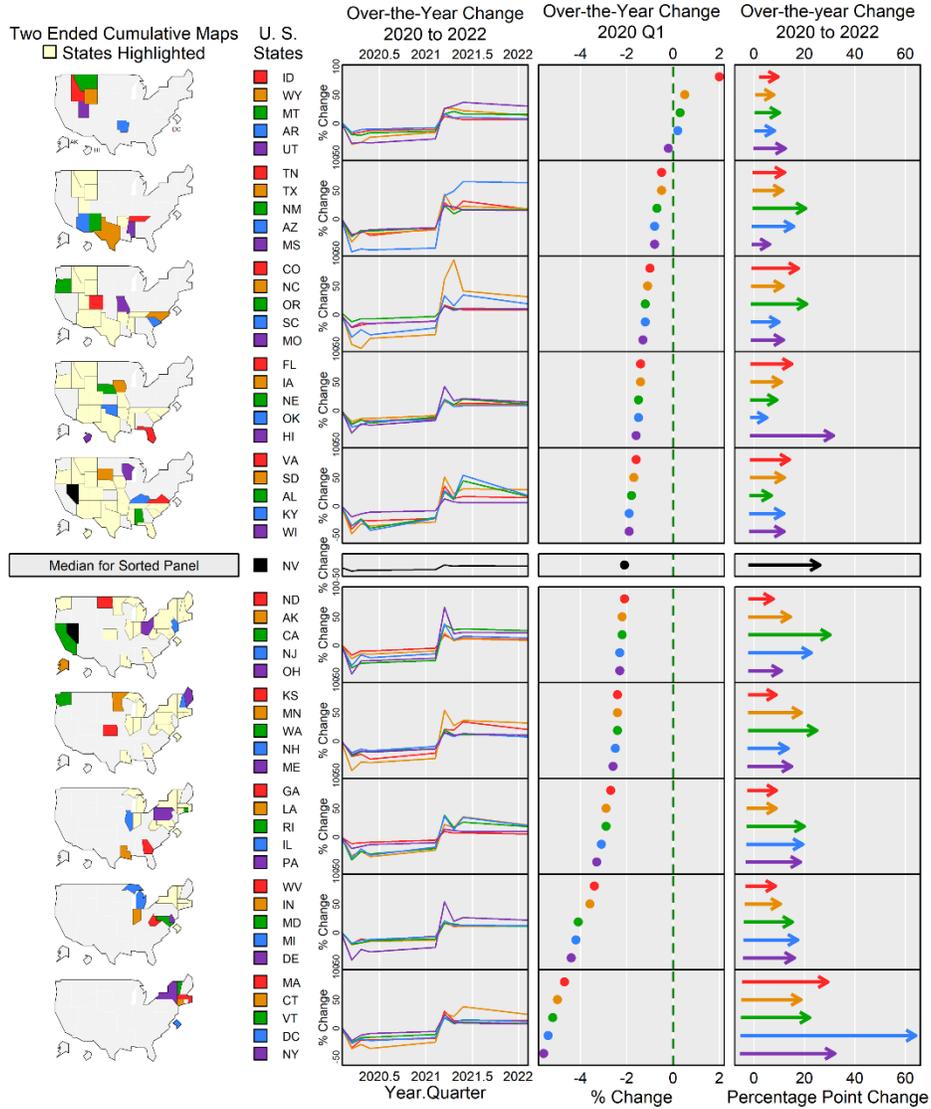

*Figure 9. The statistics are ordered by the over-the-year change in employment for the Leisure and Hospitality Industry for the change from 2019 Q1 to 2020 Q1. A time series of these one-year changes for the quarters between Q1 2020 and Q1 2022 are shown, where one can see a spike (increase) in employment in early 2021 showing some recovery in employment, which then leveled off. SOURCE: Bureau of Labor Statistics, QCEW www.bls.gov.*

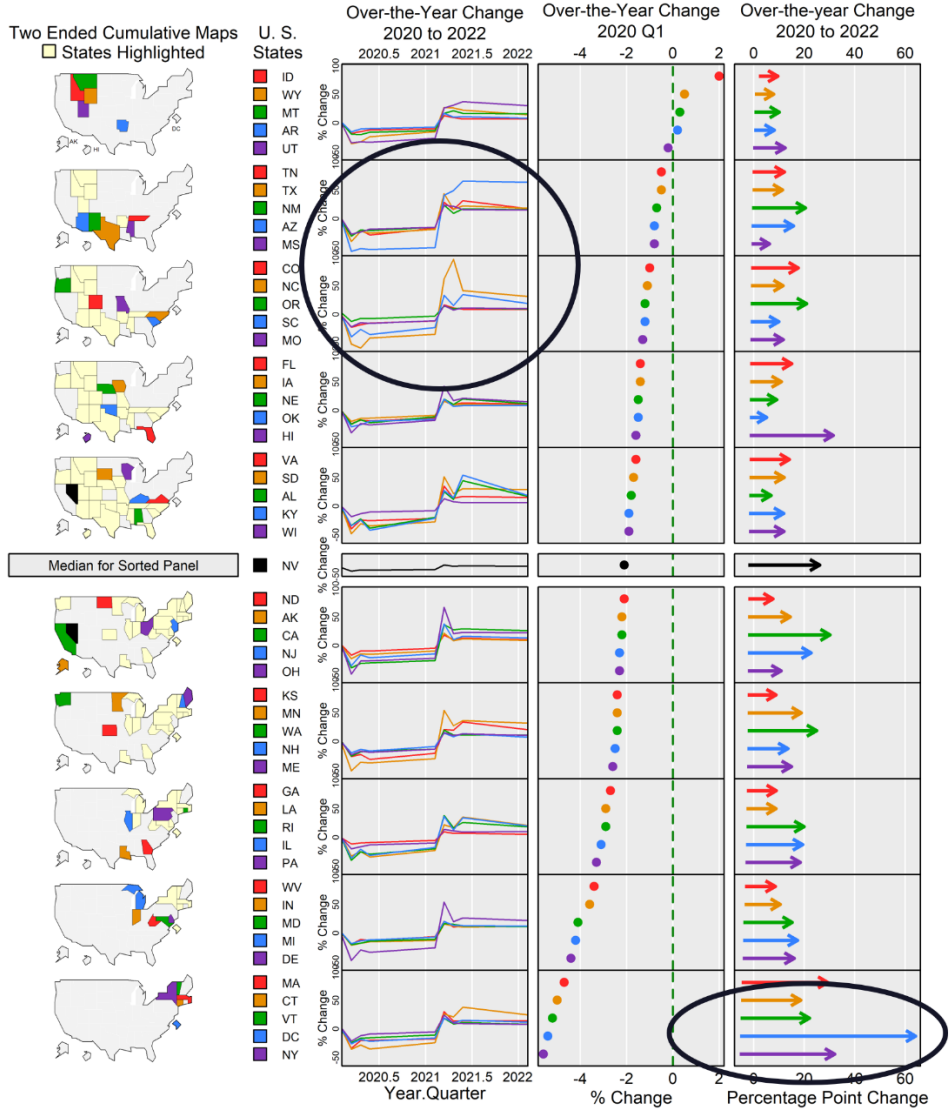

*Figure 10. This shows the same linked micromaps as in Figure 9, where some interesting changes are highlighted. Looking at the time series, Arizona and North Carolina had some extreme changes in employment during the period. From the arrows showing the percentage point change in the third column, Washington, DC had the largest change. Decision and policy makers might want to investigate these interesting phenomena. SOURCE: Bureau of Labor Statistics, QCEW www.bls.gov.*

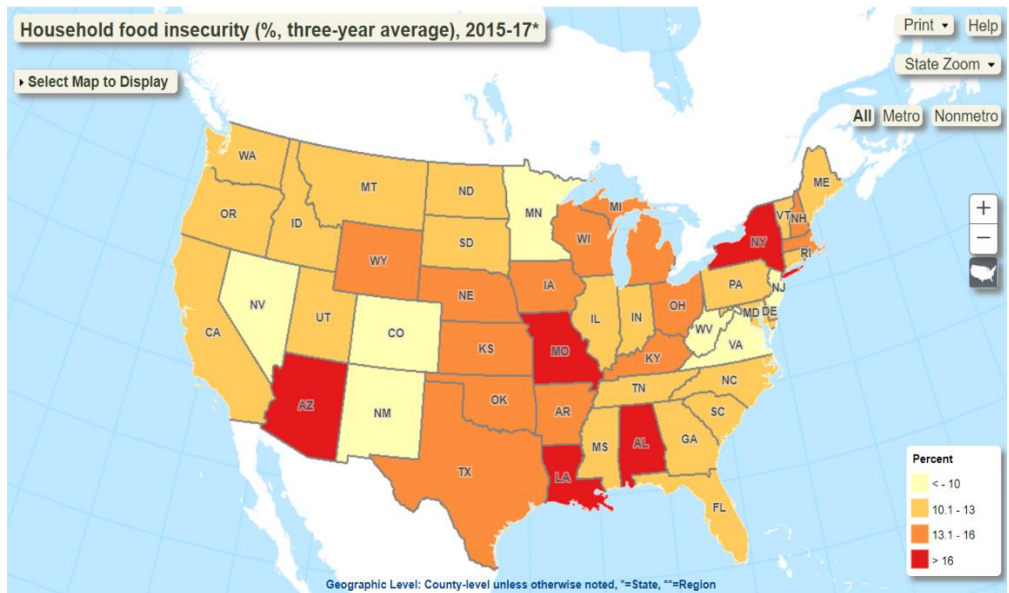

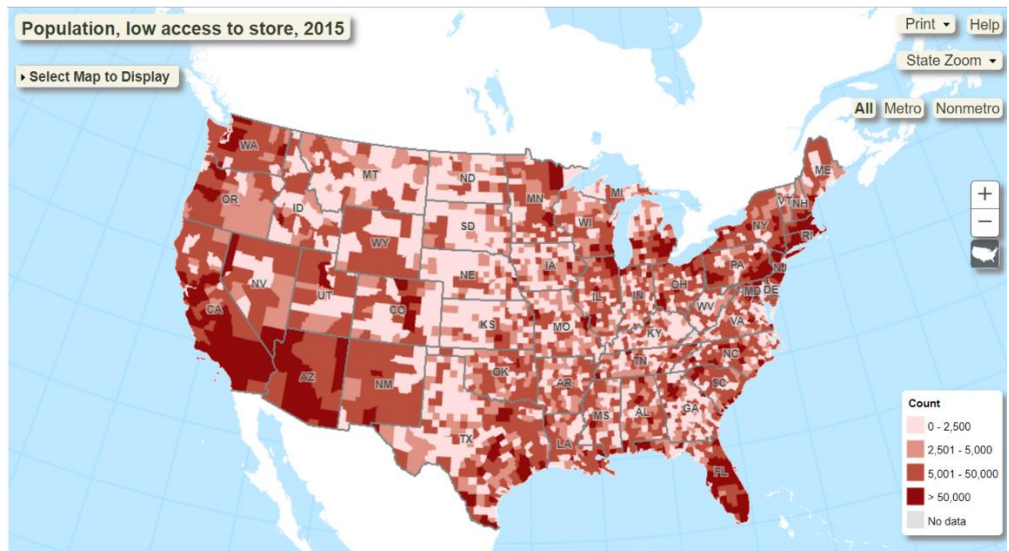

*Figure 11. Here are two examples of choropleth maps copied from the ERS Food Environment Atlas. At the top is a choropleth map at the state level showing household food insecurity, while the bottom shows a county-level choropleth map displaying the population with low access to stores. SOURCE: Economic Research Service Food Environment Atlas https://www.ers.usda.gov/data-products/food-environment-atlas/.*

## SNAP Participation and Food Insecurity

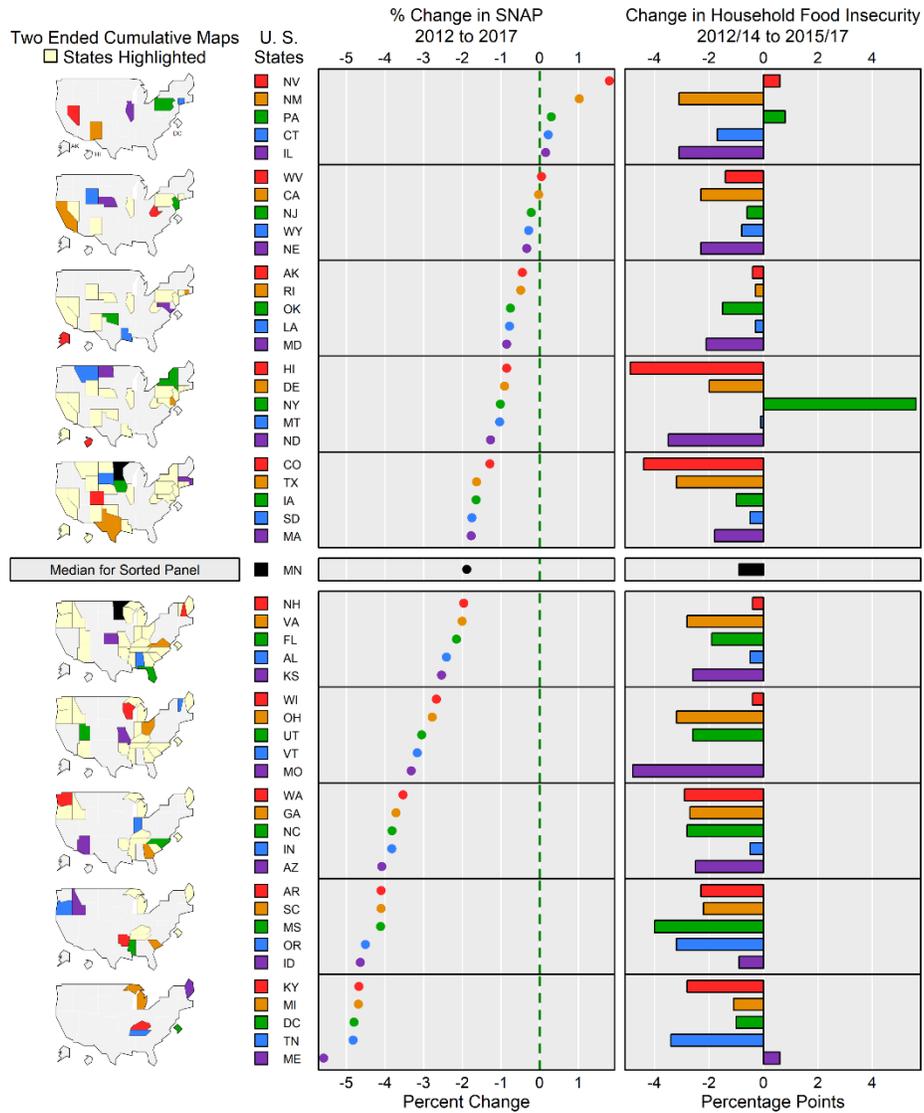

*Figure 12. This linked micromaps graphic illustrates the relationship (if any) between measures of SNAP participation and food insecurity. This displays the percent change in SNAP from 2012 to 2017 as dots (middle panel) and the change in 3-year average in household food insecurity as bars (right panel). Food insecurity for most states decreased, as did SNAP participation. Based on the bar plot, policymakers might want to look closely at New York, as that state had a large increase in reported food insecurity. Three out of the five states with increases in SNAP participation also had decreases in food insecurity, which could also be explored further by policymakers. SOURCE: Economic Research Service Food Environment Atlas https://www.ers.usda.gov/data-products/food-environment-atlas/.*

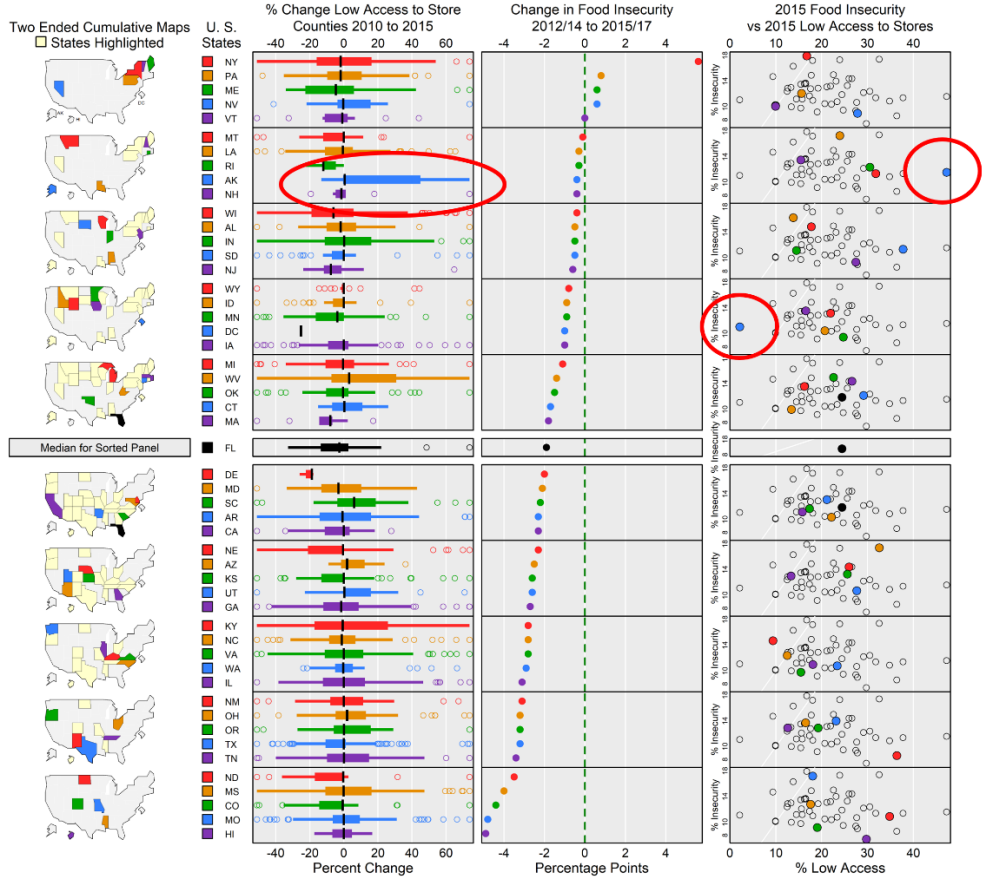

Figure 13. This graphic shows two additional ways to display statistics linked to the micromaps: boxplots and scatterplots. Policymakers might want to explore the relationship between low access to stores and food insecurity. The first data column displays boxplots of the county level change in access to stores, which indicate the distributions of the statistic, the dispersion, and the median. The last column shows a scatterplot of 2015 Food Insecurity as a function of Low Access to Stores, where there does not seem to be a strong relationship between them at the state level of aggregation. Some interesting states are circled. Alaska and Washington, DC are both extreme points in the scatterplots for access to stores, and the boxplot for Alaska is highly skewed. Decision makers might look to these areas for policy insights. SOURCE: Economic Research Service Food Environment Atlas https://www.ers.usda.gov/data-products/food-environment-atlas/.